\pdfoutput=1

\documentclass[11pt]{article}

\usepackage[final]{acl}

\usepackage{times}
\usepackage{caption}
\usepackage{subcaption}
\usepackage{latexsym}
\usepackage{graphicx}
\usepackage{listings}
\usepackage{multicol}
\usepackage{multirow}
\usepackage{footmisc}
\usepackage{footnote}
\usepackage{booktabs}
\usepackage{float}
\usepackage{tikz}
\usepackage{fontawesome}
\usepackage[T1]{fontenc}

\usepackage[utf8]{inputenc}

\usepackage{microtype}

\usepackage{inconsolata}

\usepackage{xcolor} 

\newcommand{\markllm}{\textsc{MarkLLM}}

\lstdefinestyle{github}{
    language=Python,                
    backgroundcolor=\color{white},  
    basicstyle=\ttfamily\small,     
    keywordstyle=\color{blue},      
    stringstyle=\color[rgb]{0.639,0.082,0.082}, 
    commentstyle=\color[rgb]{0,0.502,0}, 
    identifierstyle=\color{black},  
    sensitive=true,
    showstringspaces=false,         
    numberstyle=\tiny\color{gray},  
    numbers=left,                   
    numbersep=5pt,                  
    tabsize=4,                      
    breaklines=true,                
    extendedchars=true,             
    frame=single,                   
    framesep=5pt,                   
    xleftmargin=5pt,                
    xrightmargin=5pt,               
    rulecolor=\color{black},        
    captionpos=b                    
}

\lstdefinestyle{bash}{
    language=bash,
    basicstyle=\ttfamily\scriptsize,
    keywordstyle=\color{blue},
    commentstyle=\color{gray},
    showstringspaces=false,
    captionpos=b,
    frame=single,
    numbers=left,
    numberstyle=\tiny\color{gray},
    breaklines=true,
    breakatwhitespace=true,
    literate={`}{{\textquotesingle}}1
             {'}{{\textquotesingle}}1,
    morekeywords={python},
}

\lstdefinestyle{jsonStyle}{
    backgroundcolor=\color{white},
    basicstyle=\footnotesize\ttfamily,
    breaklines=true,
    captionpos=b,
    commentstyle=\color{gray},
    escapeinside={(*@}{@*)},
    keywordstyle=\color{blue},
    stringstyle=\color{red},
    frame=tb,
    numbers=left,
    numbersep=5pt,
    numberstyle=\tiny\color{gray},
    showstringspaces=false,
    tabsize=2
}

\lstnewenvironment{jsoncode}[1][]{
    \lstset{style=jsonStyle, #1}
}{}

\title{\markllm: An Open-Source Toolkit for LLM Watermarking}

\author{
Leyi Pan\textsuperscript{1},
~~~ Aiwei Liu\textsuperscript{1}\thanks{Project Leader},
~~~ Zhiwei He\textsuperscript{2},
~~~ Zitian Gao\textsuperscript{3},
~~~ Xuandong Zhao\textsuperscript{4}, \\
~~~ \bf{Yijian Lu}\textsuperscript{5},
~~~ Bingling Zhou\textsuperscript{2},
~~~ Shuliang Liu\textsuperscript{6,7},
~~~ Xuming Hu\textsuperscript{6,7},
~~~ Lijie Wen\textsuperscript{1}\thanks{Corresponding Author}, \\
~~~ \bf{Irwin King}\textsuperscript{5},
~~~ Philip S. Yu\textsuperscript{8}\\
\textsuperscript{1}Tsinghua University~~~
\textsuperscript{2}Shanghai Jiao Tong University~~~
\textsuperscript{3}The University of Sydney~~~\\
\textsuperscript{4}UC Santa Barbara~~~
\textsuperscript{5}The Chinese University of Hong Kong~~~\\
\textsuperscript{6}The Hong Kong University of Science and Technology (Guangzhou)~~~\\
\textsuperscript{7}The Hong Kong University of Science and Technology~~~
\textsuperscript{8}University of Illinois at Chicago\\
{\tt\small panly24@mails.tsinghua.edu.cn, liuaw20@mails.tsinghua.edu.cn, xuminghu@hkust-gz.edu.cn} \\
{\tt\small wenlj@tsinghua.edu.cn, king@cuhk.edu.hk, psyu@uic.edu}
}

\begin{document}
\maketitle
\begin{abstract}

Watermarking for Large Language Models (LLMs), which embeds imperceptible yet algorithmically detectable signals in model outputs to identify LLM-generated text, has become crucial in mitigating the potential misuse of LLMs. However, the abundance of LLM watermarking algorithms, their intricate mechanisms, and the complex evaluation procedures and perspectives pose challenges for researchers and the community to easily understand, implement and evaluate the latest advancements.
To address these issues, we introduce \markllm, an open-source toolkit for LLM watermarking. \textsc{MarkLLM} offers a unified and extensible framework for implementing LLM watermarking algorithms, while providing user-friendly interfaces to ensure ease of access. Furthermore, it enhances understanding by supporting automatic visualization of the underlying mechanisms of these algorithms. For evaluation, \textsc{MarkLLM} offers a comprehensive suite of 12 tools spanning three perspectives, along with two types of automated evaluation pipelines.
Through \markllm, we aim to support researchers while improving the comprehension and involvement of the general public in LLM watermarking technology, fostering consensus and driving further advancements in research and application. Our code is available at \href{https://github.com/THU-BPM/MarkLLM}{https://github.com/THU-BPM/MarkLLM}.
\end{abstract}
\section{Introduction}
\label{sec:intro}
The emergence of Large Language Models (LLMs) like ChatGPT \citep{openai2022}, GPT-4 \citep{OpenAI2023GPT4TR}, and LLaMA \citep{touvron2023llama} has significantly enhanced various tasks, including information retrieval \citep{zhu2023large}, content comprehension \citep{xiao2023evaluating}, and creative writing \citep{gomez2023confederacy}. However, in the digital era, the remarkable proficiency of LLMs in generating high-quality text has also brought several issues to the forefront, including individuals impersonation \citep{salewski2023incontext}, academic paper ghostwriting \citep{vasilatos2023howkgpt}, and the proliferation of LLM-generated fake news \citep{megias2021dissimilar}. These issues highlight the urgent need for reliable methods to distinguish between human and LLM-generated content, particularly to prevent the spread of misinformation and ensure the authenticity of digital communication. In the light of this, LLM watermarking technology \citep{DBLP:conf/icml/KirchenbauerGWK23, aronsonpowerpoint, liu2024preventing, pan2024waterseeker, liu2024can} has been developed as a promising solution. By incorporating distinct features during the text generation process, LLM outputs can be uniquely identified using specially designed detectors.

As a developing technology, LLM watermarking urgently requires consensus and support from both within and outside the field. However, due to the proliferation of watermarking algorithms, their relatively complex mechanisms, the diversity of evaluation perspectives and metrics, as well as the intricate procedure of evaluation process, significant efforts are required by both researchers and the general public to easily experiment with, comprehend, and evaluate watermarking algorithms.

\begin{figure*}[t] 
  \centering
  \includegraphics[width=\textwidth]{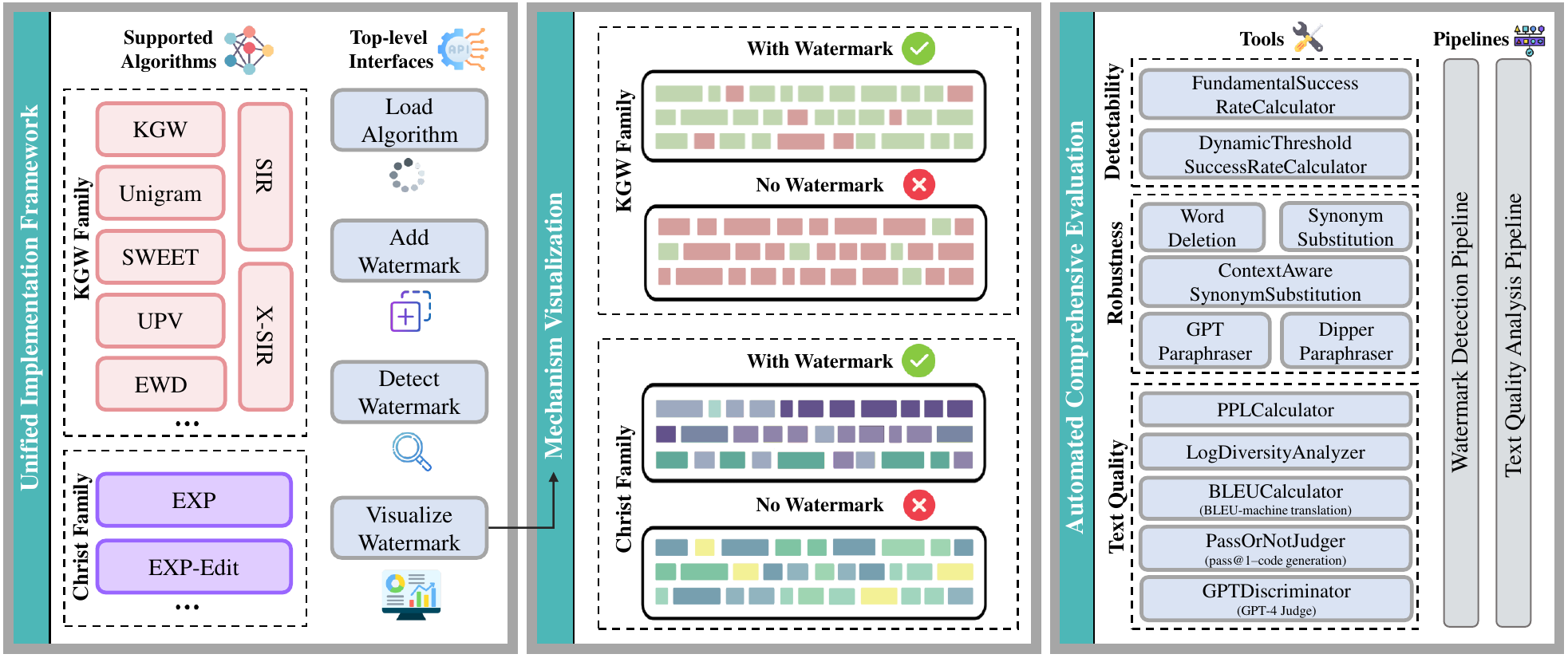} 
  \caption{Architecture overview of \textsc{MarkLLM}.}
  \label{fig:overview}
\end{figure*}

To bridge this gap, we introduce \textsc{MarkLLM}, an open-source toolkit for LLM watermarking. Figure \ref{fig:overview} overviews the architecture of \textsc{MarkLLM}.
Our main contributions are summarized as follows:

\vspace{6pt}

\noindent\textbf{1) From a Functional Perspective}:

\begin{itemize}
    \item[\faWrench] Implementation framework: \textsc{MarkLLM} offers a unified and extensible framework for implementing LLM watermarking algorithms, currently supporting nine specific algorithms from two key families: KGW \citep{DBLP:conf/icml/KirchenbauerGWK23} and Christ \citep{christ2023undetectable} family.
    \item[\faSliders] Unified top-calling interfaces: \textsc{MarkLLM} provides consistent, user-friendly interfaces for loading algorithms, producing watermarked text generated by LLMs, conducting detection processes, and gathering data necessary for visualization.
    \item[\faPaintBrush] Visualization solutions: Custom visualization solutions are provided for both major watermarking algorithm families, enabling users to visualize the mechanisms of different algorithms under various configurations with real-world examples.
    \item[\faBarChart] Evaluation module: The toolkit includes 12 evaluation tools that address three critical perspectives: detectability, robustness, and impact on text quality. It also features two types of automated evaluation pipelines that support user customization of datasets, models, evaluation metrics and attacks, facilitating flexible and comprehensive assessments.
\end{itemize}

\vspace{-3pt}

\noindent\textbf{2) From a Design Perspective}: \textsc{MarkLLM} is designed with a modular, loosely coupled architecture, ensuring its scalability and flexibility. This design choice facilitates the integration of new algorithms, the addition of innovative visualization techniques, and the extension of the evaluation toolkit by future developers.

\vspace{6pt}

\noindent\textbf{3) From an Experimental Perspective}: Utilizing \textsc{MarkLLM} as a research tool, we perform in-depth evaluations of the performances of the nine included algorithms, offering substantial insights and benchmarks that will be invaluable for ongoing and future research in LLM watermarking.

\vspace{6pt}

\begin{figure*}[t] 
  \centering
  \includegraphics[width=\textwidth]{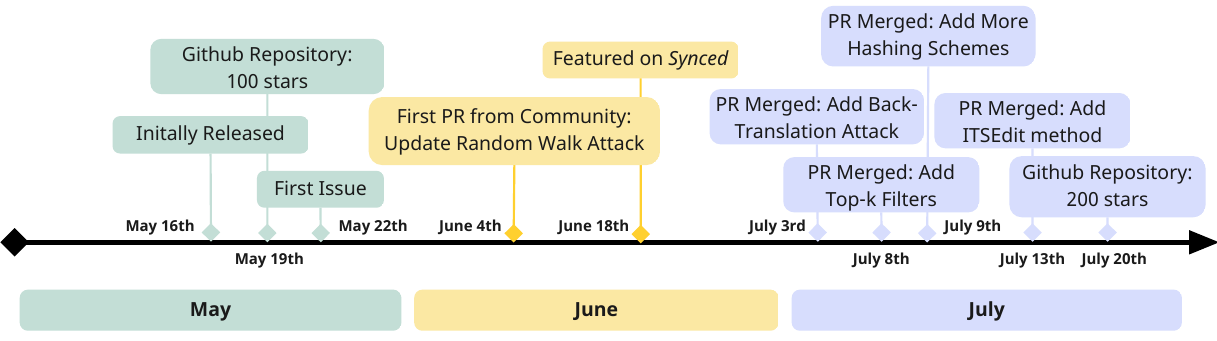} 
  \caption{Timeline of the MarkLLM ecosystem since its initial release.}
  \label{fig:timeline}
\end{figure*}

\noindent\textbf{4) From an Ecosystem Perspective}:  \textsc{MarkLLM} provides a comprehensive set of resources, including an installable Python package (a \href{https://github.com/THU-BPM/MarkLLM}{GitHub repository} and a \href{https://pypi.org/project/markllm/}{pip package}) with detailed installation and usage instructions, and an online \href{https://colab.research.google.com/drive/169MS4dY6fKNPZ7-92ETz1bAm_xyNAs0B?usp=sharing#scrollTo=sAzv2lgqG9WL}{Jupyter notebook demo} hosted on Google Colab. Since its initial release, \textsc{MarkLLM} has garnered significant attention from researchers and developers, who have actively engaged with the project through stars, forks, issues, and pull requests, fostering continuous development and improvement. Figure \ref{fig:timeline} depicts the evolution of the \textsc{MarkLLM} ecosystem since its initial release.  Due to the scope of this paper, we focus on presenting the core functionalities of \textsc{MarkLLM}, while acknowledging the broader ecosystem and community contributions that have emerged around the project.

\section{Background}
\label{sec:background}
\subsection{LLM Watermarking Algorithms}
\label{sec:background_alg}
LLM watermarking methods can be classified into the KGW Family and the Christ Family. The KGW Family modifies logits to generate watermarked output, while the Christ Family alters the sampling process.

The KGW method \citep{DBLP:conf/icml/KirchenbauerGWK23} partitions the vocabulary into green and red lists, adding bias to green list tokens during generation. A statistical metric based on the green word proportion is used for detection. Various modifications have been proposed to improve text quality \citep{hu2023unbiased,wu2023dipmark,takezawa2023necessary}, information capacity \citep{wang2023towards,yoo2023advancing,fernandez2023three}, robustness \citep{zhao2023provable,liu2024a,ren2023robust,he2024can,zhang2024watermarks}, adapt to low-entropy scenarios \citep{lee2023wrote,lu2024entropy}, and enable public detection \citep{liu2024an,cryptoeprint:2023/1661}.

\citet{christ2023undetectable} used pseudo-random numbers to guide sampling in a binary LLM. \citet{aronsonpowerpoint} developed an algorithm for real-world LLMs using EXP-sampling, where a pseudo-random sequence is generated based on previous tokens to select the next token. Watermark detection measures the correlation between the text and the sequence. \citet{kuditipudi2023robust} suggested using edit distance for robust detection.

\subsection{Evaluation Perspectives}
\label{sec:background_eva}
Evaluating watermarking algorithms involves multiple dimensions \citep{liu2023survey}:

\vspace{3pt}

\noindent\textbf{1) Watermark Detectability}: The ability to discern watermarked text from natural content.

\vspace{3pt}

\noindent\textbf{2) Robustness Against Tampering Attacks}: The watermark should withstand minor modifications and remain detectable.

\vspace{3pt}

\noindent\textbf{3) Impact on Text Quality}: Watermarking may affect the quality of generated text. This impact can be measured by perplexity, diversity, and performance in downstream tasks.
\section{\textsc{MarkLLM}}
\label{sec:markllm}
\subsection{Unified Implementation Framework}
\label{sec:markllm_invocation}

Many watermarking algorithms have been proposed, but their implementations lack standardization, leading to several issues:

\vspace{3pt}

\noindent\textbf{1) Lack of Standardization in Class Design}: Insufficiently standardized class designs make optimizing or extending existing methods difficult.

\vspace{3pt}

\noindent\textbf{2) Lack of Uniformity in Top-Level Calling Interfaces}: Inconsistent interfaces make batch processing and replicating different algorithms cumbersome and labor-intensive.

\vspace{3pt}

\noindent\textbf{3) Code Standard Issues}: Modifying settings across multiple code segments, lack of consistent documentation, hard-coded values, and inconsistent error handling complicate customization, effective use, adaptability, and debugging efforts.

\begin{figure}[h!] 
  \centering
\includegraphics[width=0.48\textwidth]{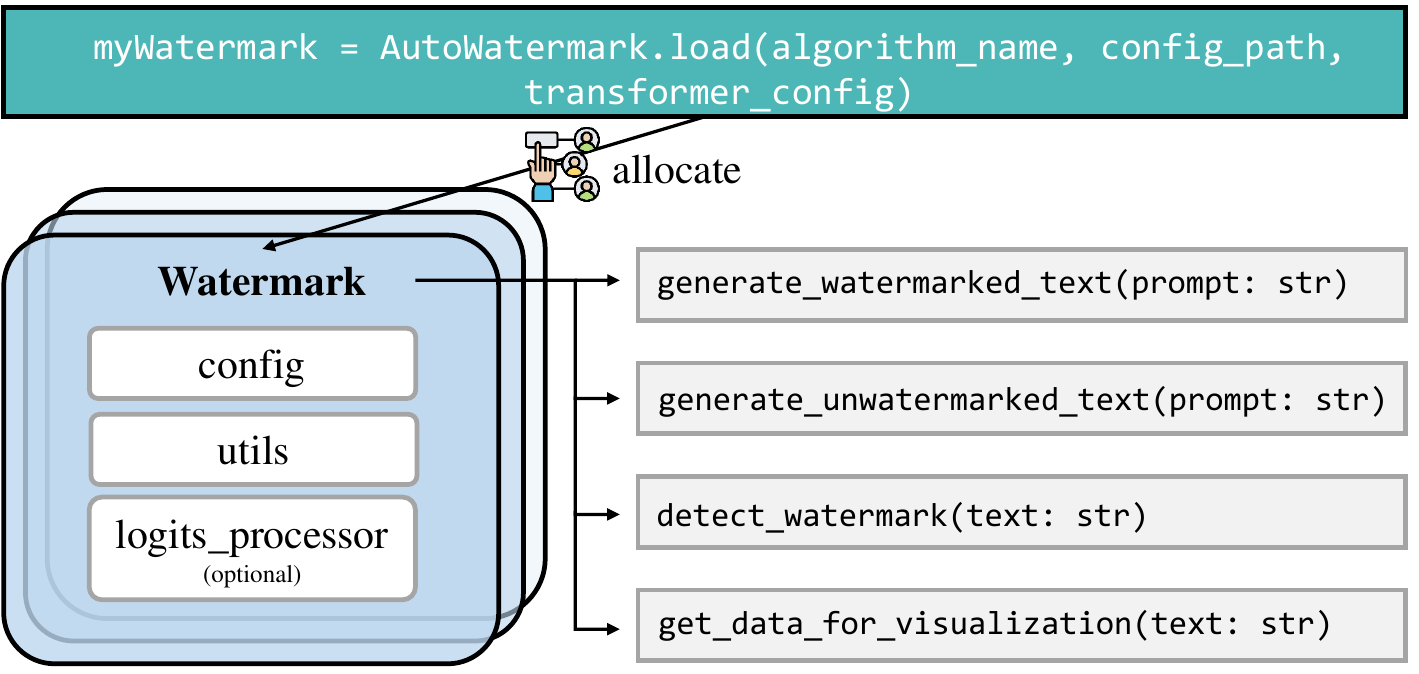} 
  \caption{Unified implementation framework of LLM watermarking algorithms.}
  \label{fig:implementation}
\end{figure}

Our toolkit offers a unified implementation framework that enables convenient invocation of various state-of-the-art algorithms under flexible configurations. Figure \ref{fig:implementation} demonstrates the design of this framework.

\noindent\textbf{AutoWatermark.} This class is responsible for algorithm allocation. Its \textit{.load()} method locates the corresponding algorithm class using \textit{algorithm\_name} and accesses its configuration\footnote{For each watermarking algorithm, all user-modifiable parameters are consolidated into a dedicated configuration file, facilitating easy modifications.} for initialization via \textit{config\_path}.

\noindent\textbf{Watermark.} Each watermarking algorithm has its own class, collectively referred to as the Watermark class. This class includes three data members: \textit{config}, \textit{utils}, and \textit{logits\_processor} (only for algorithms in the KGW Family). \textit{config} holds algorithm parameters, while \textit{utils} comprises helper functions and variables. For algorithms within the KGW family, \textit{logits\_processor} is designed to manipulate logits and is integrated into \textit{model.generate()} for processing during execution.

\begin{figure*}[t] 
  \centering
\includegraphics[width=0.86\textwidth]{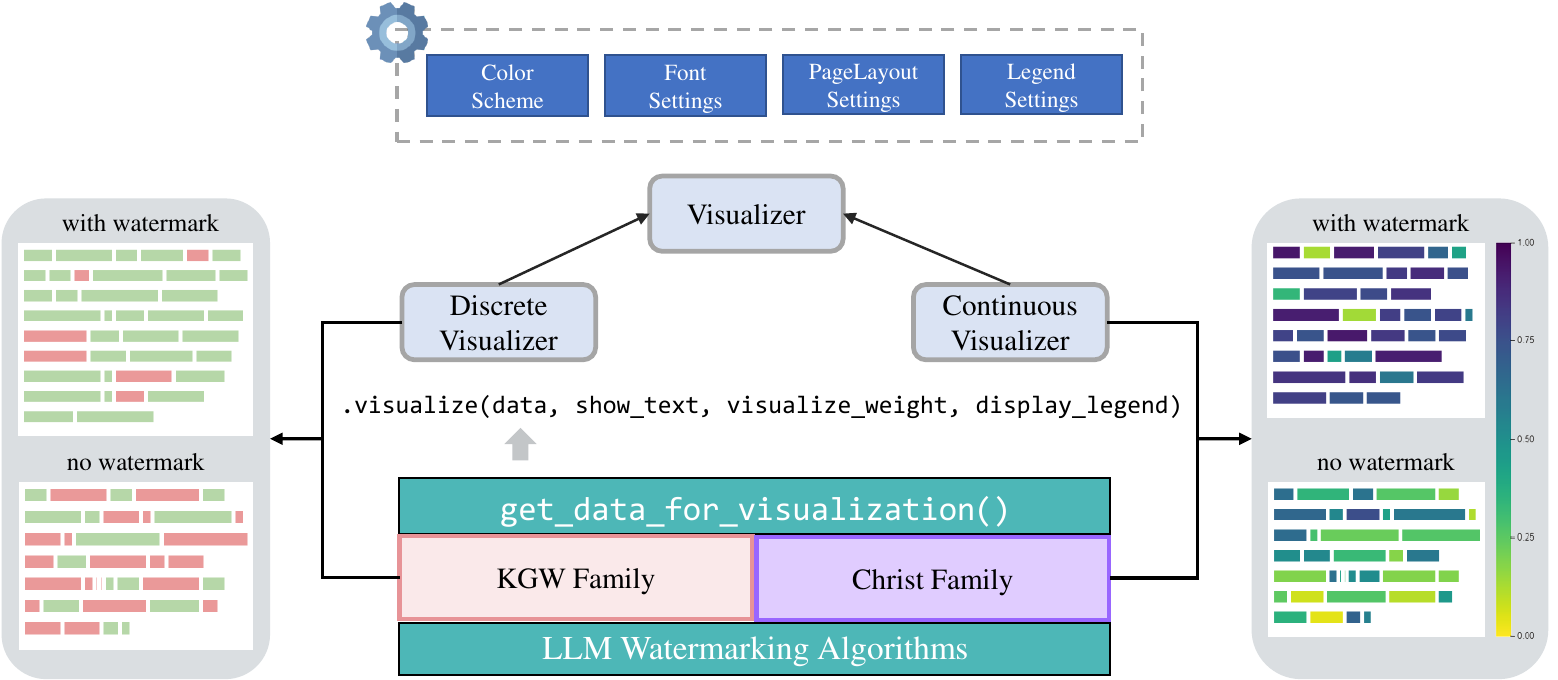} 
  \caption{Implementation framework of mechanism visualization.}
\label{fig:mechanism_visualization}
\end{figure*}

\noindent\textbf{Top-level Interfaces.} Each algorithm has four top-level interfaces for generating watermarked text, generating unwatermarked text, detecting watermarks, and obtaining data for visualization (detailed in Section \ref{sec:markllm_mechanism}). The framework's distributive design using an AutoWatermark class allows developers to easily add interfaces to any algorithm class without impacting others.

\subsection{Mechanism Visualization}
\label{sec:markllm_mechanism}

To improve understanding of the mechanisms used by different watermark algorithms, we have developed a visualization module that provides tailored visualization solutions for the two algorithm families.

\subsubsection{Visualization Solutions}

\textbf{KGW Family}. As detailed in Section \ref{sec:background_alg}, KGW family algorithms manipulate LLM output logits to prefer green tokens over red ones and employ statistical methods for detection. Our visualization technique clearly highlights red and green tokens in the text, offering insights into the token-level detection results.

\vspace{3pt}

\noindent\textbf{Christ Family}. Algorithms within Christ family involves guiding each token selection using a pseudo-random sequence and detect watermarks by calculating the correlation between the sequence and the text. To visualize this mechanism, we use a color gradient to represent the alignment value of each token and the pseudo-random sequence, where darker shades indicate stronger alignment.

\subsubsection{Architecture Design}
This section offers a detailed description of the architectural frameworks essential for the effective implementation of the aforementioned visualization strategies. Figure \ref{fig:mechanism_visualization} demonstrates the implementation framework of mechanism visualization.

\vspace{3pt}

\noindent\textbf{get\_data\_for\_visualization}: This interface, defined for each algorithm, returns a VisualizationData object containing \textit{decoded\_tokens} and \textit{highlight\_value}. For the KGW family, \textit{highlight\_value} is one-hot, differentiating red and green tokens; for the Christ family, it represents a continuous correlation value.

\vspace{3pt}

\noindent\textbf{Visualizer}: It initializes with a VisualizationData object and performs visualization via the \textit{.visualize()} method, with subclasses overriding approach to implement specific visualizations.

\vspace{3pt}

\noindent\textbf{DiscreetVisualizer}: Tailored for KGW family algorithms, it uses red/green highlight values to color-code text based on values.

\vspace{3pt} 

\noindent\textbf{ContinuousVisualizer}: Tailored for Christ family algorithms, it highlights tokens using a [0,1] color scale based on their alignment with pseudo-random numbers.

\vspace{3pt}

\noindent\textbf{Flexible Visualization Settings}: Our Visualizer supports multiple configurable options for tailored visualizations, including ColorScheme, FontSettings, PageLayoutSettings, and LegendSetting, allowing for extensive customization.

\vspace{3pt}

\subsubsection{Visualization Result}
\textbf{KGW Family}. The leftmost part of Figure \ref{fig:mechanism_visualization} shows that in the text with watermarks, there is a relatively high proportion of green tokens. The z-score, a statistical measure, is defined as:
\[
    z = \frac{|s|_G - \gamma T}{\sqrt{T\gamma(1-\gamma)}} 
\]
where $|s|_G$ is the number of green tokens, $T$ is the total number of tokens, and $\gamma$ is the proportion of the green token list in partitioning (0.5 in this case). The z-score for `text with watermark' is notably higher than that for `text without watermark'. Setting a reasonable z-score threshold can effectively distinguish between the two.

\vspace{3pt}

\noindent\textbf{Christ Family}. As depicted in the rightmost part of Figure \ref{fig:mechanism_visualization}, it is noticeable that tokens within text containing watermarks generally exhibit darker hues compared to those without, indicating a higher influence of the sequence during the generation process on the former. 
\subsection{Automated Comprehensive Evaluation}
\label{sec:markllm_eva}
Evaluating an LLM watermarking algorithm is complex, as it involves considering multiple perspectives, such as watermark detectability, robustness against tampering, and impact on text quality (see Section \ref{sec:background_eva}). Each perspective may require different metrics, attack scenarios, and tasks. The evaluation process typically includes steps like model and dataset selection, watermarked text generation, post-processing, watermark detection, text tampering, and metric computation.

To simplify the evaluation process, \textsc{MarkLLM} offers twelve user-friendly tools, including metric calculators and attackers, covering the three main evaluation perspectives. Additionally, \textsc{MarkLLM} provides two types of customizable automated demo pipelines, allowing for easy configuration and use.

\begin{table}[t]
\caption{Evaluation Tools in MarkLLM.}
\centering
\begin{tabular}{p{2cm} p{5cm}}
\hline
\small Perspective & \small Tools \\ \hline
\multirow{2}{*}{\small Detectability} & \small FundamentalSuccessRateCalculator \\
& \small DynamicThresholdSuccessRateCalculator \\ 
\midrule
\multirow{5}{*}{\small Robustness} & \small WordDeletion \\
& \small SynonymSubstitution \\ 
& \small ContextAwareSynonymSubstitution \\ 
& \small GPTParaphraser \\ 
& \small DipperParaphraser \\ 
\midrule
\multirow{5}{*}{\small Text Quality} & \small PPLCaluclator \\
& \small LogDiversityAnalyzer \\ 
& \small BLEUCalculator \\ 
& \small PassOrNotJudger \\ 
& \small GPTDiscriminator \\ 
\hline
\end{tabular}
\label{tab:tools}
\end{table}

\textsc{MarkLLM} provides a comprehensive set of tools for evaluating LLM watermarking algorithms, as summarized in Table \ref{tab:tools}. These tools cover detectability, including success rate calculators with fixed and dynamic thresholds; robustness, featuring word-level and document-level text tampering attacks using WordNet \citep{miller1995wordnet}, BERT \citep{devlin2018bert}, OpenAI API, and the Dipper model \citep{krishna2023paraphrasing}; and text quality, assessing fluency, variability, and performance on downstream tasks using perplexity, diversity, BLEU, pass-or-not judger, and GPT discriminator with GPT-4 \citep{OpenAI2023GPT4TR}.

\vspace{3pt}

\begin{figure}[t]
    \centering
    \includegraphics[width=0.48\textwidth]{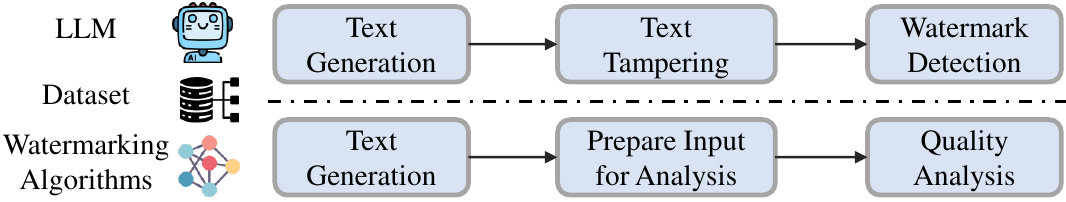}
    \caption{The standardized process of evaluation pipelines, the upper for watermark detection pipeline, and the lower for text quality analysis pipeline.}
    \label{fig:pipeline}
\end{figure}

\textbf{Evaluation Pipelines.} \textsc{MarkLLM} provides two evaluation pipelines: one for assessing watermark detectability with and without attacks, and another for analyzing the impact of these algorithms on text quality.

The upper part of Figure \ref{fig:pipeline} shows the standardized process of watermark detection. We have implemented two pipelines: \textbf{WMDetect} for watermarked text detection and \textbf{UWMDetect} for unwatermarked text detection. The lower part of Figure \ref{fig:pipeline} illustrates the unified process of text quality analysis. Pairs of watermarked and unwatermarked texts are generated and fed into a designated text quality analyzer to produce detailed analysis and comparison results. We have implemented three pipelines for different evaluation scenarios:

\textbf{DirectQual}. This pipeline directly compares the characteristics of watermarked and unwatermarked texts using metrics such as perplexity (PPL) and log diversity.

\textbf{RefQual}. This pipeline evaluates text quality by comparing both watermarked and unwatermarked texts with a common reference text. It is ideal for scenarios that require specific downstream tasks, such as machine translation and code generation.

\textbf{ExDisQual}. This pipeline employs an external judger, such as GPT-4 \citep{OpenAI2023GPT4TR}, to assess the quality of both watermarked and unwatermarked texts based on user-provided task descriptions. This method is valuable for advanced, AI-based analysis of the subtle effects of watermarking.

\section{User Examples}
The following code snippets demonstrate examples of how to use MarkLLM in one's project. For more real cases, please see the \href{https://www.youtube.com/watch?v=QN3BhNvw14E}{demo video}.
\subsection{Watermarking Algorithm Invocation}
\label{sec:appendix_examples_alg}
\begin{lstlisting}[style=github]
# Load algorithm
myWatermark = AutoWatermark.load('KGW', 'config/KGW.json', transformers_config)
# Generate watermarked text
watermarked_text = myWatermark.generate_watermarked_text(prompt)
# Detect watermark
detect_result = myWatermark.detect_watermark(watermarked_text)
\end{lstlisting}

\subsection{Mechanism Visualization}
\label{sec:appendix_examples_visual}
\begin{lstlisting}[style=github]
# Get data for visualization
watermarked_data = myWatermark.get_data_for_visualization(watermarked_text)
# Init visualizer
visualizer = DiscreetVisualizer(ColorSchemeForDiscreetVisualization(), FontSettings(), PageLayoutSettings(), DiscreetLegendSettings())
# Visualize
watermarked_img = visualizer.visualize(watermarked_data)
\end{lstlisting}

\subsection{Evaluation Pipelines Invocation}
\label{sec:appendix_examples_eva}
\begin{lstlisting}[style=github]
# Dataset
my_dataset = C4Dataset('dataset/c4/processed_c4.json')
# WMDetect
pipeline1 = WatermarkedTextDetectionPipeline(my_dataset) 
# UWMDetect
pipeline2 = UnWatermarkedTextDetectionPipeline(dataset=my_dataset)
# Init calculator
calculator = DynamicThresholdSuccessRateCalculator(labels=['TPR', 'F1'], rule='best')
# Calculate success rate
print(calculator.calculate(pipeline1.evaluate(my_watermark), pipeline2.evaluate(my_watermark)))
\end{lstlisting}

\section{Experiment}
\label{sec:experiment}
Using \textsc{MarkLLM} as a research tool, we conduct evaluations on nine watermarking algorithms, assessing their detectability, robustness, and impact on text quality. Our experiments demonstrate that \textsc{MarkLLM} can reproduce the results of previous experiments with low cost through simple scripts. For details on the experimental setup and the obtained results, please refer to Appendix \ref{appendix:experiment}.
\section{Conclusion}
\textsc{MarkLLM} is a comprehensive open-source toolkit for LLM watermarking. It allows users to easily try various state-of-the-art algorithms with flexible configurations to watermark their own text and conduct detection, and provides clear visualizations to gain insights into the underlying mechanisms. The inclusion of convenient evaluation tools and customizable evaluation pipelines enables automatic and thorough assessments from various perspectives. As LLM watermarking evolves, \textsc{MarkLLM} aims to be a collaborative platform that grows with the research community. By providing a solid foundation and inviting contributions, we aim to foster a vibrant ecosystem where researchers and developers can work together to advance the state-of-the-art in LLM watermarking technology.
\section*{Limitations}
MarkLLM is a comprehensive toolkit for implementing, visualizing, and evaluating LLM watermarking algorithms. However, it currently only integrates a subset of existing methods and does not yet support some recent approaches that directly embed watermarks into model parameters during training \citep{xu2024learning,gu2024learnability}. We anticipate future contributions to expand MarkLLM's coverage and enhance its versatility.

In terms of visualization, we have provided one tailored solution for each of the two main watermarking algorithm families. While these solutions offer valuable insights, there is room for more creative and diverse visualization designs.

Regarding evaluation, we have covered aspects such as detectability, robustness, and text quality impact. However, our current toolkit may not encompass all possible scenarios, such as spoofing attack and CWRA \citep{he2024can}.

We acknowledge that \textsc{MarkLLM} has room for improvement. We warmly welcome developers and researchers to contribute their code and insights to help build a more comprehensive ecosystem for LLM watermarking. Through collaborative efforts, we can further advance this technology and unlock its full potential.

\section*{Acknowledgements}
\textsc{MarkLLM}, as an open-source toolkit, has greatly benefited from the community's feedback and contributions. We extend our sincere gratitude to all users who have raised issues on GitHub, thereby helping us improve this project. Special thanks go to Hanlin Zhang, Sheng Guan, Yiming Liu, Yichen Di, and Kai Shi for their valuable pull requests, which have significantly enhanced MarkLLM's functionality. Furthermore, we are deeply appreciative of the insightful comments provided by the reviewers and area chair, which have been instrumental in refining both our paper and the project.

\bibliography{custom}
\appendix
\newpage
\clearpage

\section{Experiment Details}
\label{appendix:experiment}
\begin{table*}[h!]
\caption{
The evaluation results of assessing the detectability of nine algorithms supported in MarkLLM. 200 watermarked texts are generated, while 200 non-watermarked texts serve as negative examples. We furnish TPR and F1-score under dynamic threshold adjustments for 10\% and 1\% FPR, alongside TPR, TNR, FPR, FNR, P, R, F1, ACC at optimal performance.
}
\centering
\resizebox{0.85\textwidth}{!}{
\begin{tabular}{lcc|cc|cccccccc}
\toprule
\multirow{2}{*}{Method} & \multicolumn{2}{c}{10\%FPR} & \multicolumn{2}{c}{1\%FPR} & \multicolumn{8}{c}{Best}\\
\cmidrule(r){2-13}
    & TPR & F1 & TPR & F1 & TPR & TNR & FPR & FNR & P & R & F1 & ACC \\
\midrule
KGW & \textbf{1.000} & 0.952 & \textbf{1.000} & 0.995 & \textbf{1.000} & \textbf{1.000} & \textbf{0.000} & \textbf{0.000} & \textbf{1.000} & \textbf{1.000} & \textbf{1.000} & \textbf{1.000} \\
Unigram & \textbf{1.000} & \textbf{0.957} & \textbf{1.000} & \textbf{0.995} & \textbf{1.000} & \textbf{1.000} & \textbf{0.000} & \textbf{0.000} & \textbf{1.000} & \textbf{1.000} & \textbf{1.000} & \textbf{1.000} \\
SWEET & \textbf{1.000} & 0.952 & \textbf{1.000} & \textbf{0.995} & \textbf{1.000} & \textbf{1.000} & \textbf{0.000} & \textbf{0.000} &  \textbf{1.000} & \textbf{1.000} & \textbf{1.000} & \textbf{1.000}  \\
UPV & $\times$ & $\times$ & $\times$ & $\times$ & \textbf{1.000} & 0.990 & 0.010 & \textbf{0.000} & 0.990 & \textbf{1.000} & 0.995 & 0.995 \\
EWD & \textbf{1.000} & 0.952 & \textbf{1.000} & \textbf{0.995} & 0.995 & \textbf{1.000} & \textbf{0.000} & 0.005 & \textbf{1.000} & 0.995 & 0.997 & 0.998 \\
SIR & 0.995 & 0.950 & 0.990 & 0.990 & 0.990 & 0.995 & 0.005 & 0.010 & 0.995 & 0.990 & 0.992 & 0.993 \\
X-SIR & 0.995 & 0.950 & 0.940 & 0.964 & 0.970 & 0.970 & 0.030 & 0.030 & 0.970 & 0.970 & 0.970 & 0.970 \\
EXP & \textbf{1.000} & 0.952 & \textbf{1.000} & \textbf{0.995} & \textbf{1.000} & \textbf{1.000} & \textbf{0.000} & \textbf{0.000} & \textbf{1.000} & \textbf{1.000} & \textbf{1.000} & \textbf{1.000} \\
EXP-Edit & \textbf{1.000} & 0.952 & 0.995 & 0.990 & 0.995 & 0.985 & 0.015 & 0.005 & 0.985 & 0.995 & 0.990 & 0.990 \\
\bottomrule
\end{tabular}
}
\vspace{1em}

\label{tab:detectability_llama}
\end{table*}

\begin{table*}[h!]
\caption{
The evaluation results of assessing the robustness of nine algorithms supported in MarkLLM. For each attack, 200 watermarked texts are generated and subsequently tampered, with an additional 200 non-watermarked texts serving as negative examples. We report the TPR and F1-score at optimal performance under each circumstance.
}
\vspace{10pt}
\centering
\resizebox{0.85\textwidth}{!}{
\begin{tabular}{lcc|cc|cc|cc|cc|cc}
\toprule
\multirow{2}{*}{Method} & \multicolumn{2}{c}{No Attack} & \multicolumn{2}{c}{Word-D} & \multicolumn{2}{c}{Word-S} & \multicolumn{2}{c}{Word-S (Context)} & \multicolumn{2}{c}{Doc-P (GPT-3.5)} & \multicolumn{2}{c}{Doc-P (Dipper)} \\
\cmidrule(r){2-13}
& TPR & F1 & TPR & F1 & TPR & F1 & TPR & F1 & TPR & F1 & TPR & F1 \\
\midrule
KGW & \textbf{1.000} & \textbf{1.000} & 0.980 & 0.985 & 0.920 & 0.915 & 0.965 & 0.958 & 0.835 & 0.803 & 0.860 & 0.785 \\
Unigram & \textbf{1.000} & \textbf{1.000} & \textbf{1.000} & \textbf{1.000} & \textbf{0.990} & 
\textbf{0.990} & \textbf{0.990} & \textbf{0.990} & \textbf{0.901} & \textbf{0.932} & 0.875 & \textbf{0.908} \\
SWEET & \textbf{1.000} & \textbf{1.000} & 0.970 & 0.975 & 0.935 & 0.903 & 0.985 & 0.980 & 0.845 & 0.813 & 0.830 & 0.779 \\
UPV & \textbf{1.000} & 0.995 & 0.970 & 0.980 & 0.885 & 0.896 & 0.985 & 0.961 & 0.830 & 0.827 & 0.862 & 0.864 \\
EWD & 0.995 & 0.997 & 0.980 & 0.982 & 0.930 & 0.921 & 0.950 & 0.955 & 0.852 & 0.825 & 0.845 & 0.784 \\
SIR & 0.990 & 0.992 & 0.950 & 0.970 & 0.945 & 0.940 & 0.960 & 0.948 & 0.891 & 0.923 & \textbf{0.894} & 0.902 \\
X-SIR & 0.970 & 0.970 & 0.940 & 0.957 & 0.910 & 0.908 & 0.895 & 0.925 & 0.875 & 0.891 & 0.835 & 0.869 \\
EXP & \textbf{1.000} & \textbf{1.000} & 0.975 & 0.980 & 0.945 & 0.950 & 0.980 & 0.985 & 0.763 & 0.772 & 0.740 & 0.793 \\
EXP-Edit & 0.995 & 0.990 & 0.995 & 0.993 & 0.983 & 0.972 & 0.990 & 0.985 & 0.872 & 0.886 & 0.845 & 0.861 \\

\bottomrule
\end{tabular}
}
\vspace{1em}

\label{tab:robustness_llama}
\end{table*}
\begin{table*}[h!]
\caption{
The evaluation results of assessing the text quality impact of the nine algorithms supported in MarkLLM. We compared 200 watermarked texts with 200 non-watermarked texts. However, due to dataset constraints, only 100 watermarked texts were compared with 100 non-watermarked texts for code generation.
}
\vspace{10pt}
\centering
\resizebox{0.96\textwidth}{!}{
\begin{tabular}{lcc|cc|c}
\toprule
\multirow{3}{*}{Method} & \multicolumn{2}{c}{Direct Analysis} & \multicolumn{2}{c}{Referenced Analysis} & External Discriminator \\
\cmidrule(r){2-6}
& \multirow{2}{*}{PPL(Ori.= 8.243)} & \multirow{2}{*}{Log Diversity(Ori.=8.517)} & Machine Translation & Code Generation & Machine Translation\\
& & & BLEU(Ori.=31.807) & pass@1(Ori.= 43.0) & GPT-4 Judge (Wat. Win Rate)\\
\midrule
KGW & 13.551 $\uparrow$ & 7.989 $\downarrow$  & 28.242 $\downarrow$ & 34.0 $\downarrow$ & 0.31 \\
Unigram & 13.723 $\uparrow$ & 7.242 $\downarrow$ & 26.075 $\downarrow$ & 32.0 $\downarrow$ & \textbf{0.33} \\
SWEET & 13.747 $\uparrow$ & 8.086 $\downarrow$ & 28.242 $\downarrow$ & \textbf{37.0} $\downarrow$ & 0.31 \\
UPV & \textbf{10.574} $\uparrow$ & 7.698 $\downarrow$ & 28.270 $\downarrow$ & \textbf{37.0} $\downarrow$ & 0.31 \\
EWD & 13.402 $\uparrow$ & 8.220 $\downarrow$ & 28.242 $\downarrow$ & 34.0 $\downarrow$ & 0.30\\
SIR & 13.918 $\uparrow$ & 7.990 $\downarrow$ & \textbf{28.830} $\downarrow$ & \textbf{37.0} $\downarrow$ & 0.31 \\
X-SIR & 12.885 $\uparrow$ & 7.930 $\downarrow$ & 28.161 $\downarrow$ & 36.0 $\downarrow$ & \textbf{0.33} \\
EXP & 19.597 $\uparrow$ & 8.187 $\downarrow$ & $\times$ & 20.0 $\downarrow$ & $\times$ \\
EXP-Edit & 21.591 $\uparrow$ & \textbf{9.046} $\uparrow$ & $\times$ & 14.0 $\downarrow$ & $\times$\\

\bottomrule
\end{tabular}
}
\vspace{1em}

\label{tab:quality_llama}
\end{table*}

\subsection{Experiment Settings}
\textbf{Dateset and Prompt}. For general-purpose text generation scenarios, we utilize the C4 dataset \citep{raffel2020exploring}. Specifically, the first 30 tokens of texts serve as prompts for generating the subsequent 200 tokens, with the original C4 texts acting as non-watermarked examples. For specific downstream tasks, we employ the WMT16 \citep{bojar-EtAl:2016:WMT1} German-English dataset for machine translation, and HumanEval \citep{chen2021evaluating} for code generation.

\vspace{3pt}

\noindent\textbf{Language Model}. For general-purpose text generation scenarios, we utilize Llama-7b \citep{touvron2023llama} as language model. For specific downstream tasks, we utilize NLLB-200-distilled-600M \citep{costa2022no} for machine translation and Starcoder \citep{li2023starcoder} for code generation. 

\vspace{3pt}

\noindent\textbf{Metrics and Attacks}. Dynamic threshold adjustment is employed to evaluate watermark detectability, with three settings provided: under a target FPR of 10\%, under a target FPR of 1\%, and under conditions for optimal F1 score performance. To assess robustness, we utilize all text tampering attacks listed in Table \ref{tab:tools}. For evaluating the impact on text quality, our metrics include PPL, log diversity, BLEU (for machine translation), pass@1 (for code generation), and assessments using GPT-4 Judge \citep{tu2023waterbench}.

\subsection{Results and Analysis} 
The results\footnote{(1) The evaluation results for UPV are only shown in the ``best" column because its watermark detection uses direct binary classification without thresholds. (2) Current implementations of Christ family algorithms are designed for decoder-only LLMs. As machine translation mainly uses encoder-decoder models, we did not report the text quality produced by EXP and EXP-edit in machine translation.} in Table \ref{tab:detectability_llama}, Table \ref{tab:robustness_llama}, and Table \ref{tab:quality_llama} demonstrate that by using the implementations of different algorithms and the evaluation pipelines provided in \textsc{MarkLLM}, researchers can effectively reproduce the experimental results from previous watermarking papers. These experiments can be conducted by running simple scripts which are accessible within the Github repository under the directory \textit{ evaluation/examples/}. The execution command can be found in Listing \ref{lst:shellscript_detect}, Listing \ref{lst:shellscript_robustness} and Listing \ref{lst:shellscript_quality}, showcasing \textsc{MarkLLM}'s capability for easy evaluation of watermark algorithms in various scenarios. 

\newpage

\begin{lstlisting}[style=bash, caption={Execution command for assessing detectability.}, label={lst:shellscript_detect}]
python evaluation/examples/assess_detectability.py --algorithm KGW --labels TPR F1 --rules target_fpr --target_fpr 0.01

python evaluation/examples/assess_detectability.py --algorithm KGW --labels TPR TNR FPR FNR P R F1 ACC --rules best

\end{lstlisting}

\begin{lstlisting}[style=bash, caption={Execution command for assessing robustness.}, label={lst:shellscript_robustness}]
python evaluation/examples/assess_robustness.py --algorithm KGW --attack `Word-D'

python evaluation/examples/assess_robustness.py --algorithm Unigram --attack `Doc-P(GPT-3.5)'
\end{lstlisting}

\begin{lstlisting}[style=bash, caption={Execution command for assessing text quality.}, label={lst:shellscript_quality}]
python evaluation/examples/assess_quality.py --algorithm KGW --metric PPL

python evaluation/examples/assess_quality.py --algorithm SIR --metric `Log Diversity'
\end{lstlisting}

\section{Comparison with Competitors}
\label{sec:appendix_comparison}
As LLM watermarking technology advances, frameworks dedicated to this field have emerged. WaterBench \citep{tu2023waterbench} and Mark My Words \citep{piet2023mark} are two prominent examples. WaterBench focuses on assessing the impact of KGW \citep{DBLP:conf/icml/KirchenbauerGWK23}, Unigram \citep{zhao2023provable}, and KGW-v2 \citep{kirchenbauer2023reliability} on text quality, while Mark My Words evaluates the performance of KGW, EXP \citep{aronsonpowerpoint}, Christ \citep{christ2023undetectable}, and EXP-Edit \citep{kuditipudi2023robust} across text quality, robustness against tampering, and number of tokens needed for detection.

While these frameworks primarily focus on benchmark construction, similar to the evaluation module in \textsc{MarkLLM}, \textsc{MarkLLM} distinguishes itself as the first comprehensive multi-functional toolkit. It offers easy-to-use evaluation tools and automated pipelines that cover the aforementioned assessment perspectives, and also provides a unified implementation framework for watermarking algorithms and visualization tools for their underlying mechanisms. This enhances its utility and versatility. The integration of these functionalities makes \textsc{MarkLLM} a more accessible resource, enabling convenient usage, understanding, evaluation, and selection of diverse watermarking algorithms by researchers and the broader community. This plays a crucial role in fostering consensus both within and beyond the field.
\end{document}